\documentstyle[aps,prl,epsf,multicol]{revtex}

\newcommand{\figurewidth}{80mm}

\begin{document}

\draft

\title{Criticality in one dimension with inverse square-law potentials}

\author{Erik Luijten$^{1,}$\thanks{Electronic address: luijten@ipst.umd.edu}
and Holger Me{\ss}ingfeld$^{2,}$\thanks{Presently at Accenture, 246
Kifissias Avenue, 15231 Halandri, Athens, Greece.}}

\address{$^1$Institute for Physical Science and Technology, University of
 Maryland, College Park, MD 20742, USA}

\address{$^2$Institut f\"ur Physik, WA 331, Johannes Gutenberg-Universit\"at,
 D-55099 Mainz, Germany}

\date{February 2, 2001}

\maketitle

\begin{abstract}
  It is demonstrated that the scaled order parameter for ferromagnetic Ising
  and three-state Potts chains with inverse-square interactions exhibits a {\em
  universal\/} critical jump, in analogy with the superfluid density in helium
  films. Renormalization-group arguments are combined with numerical
  simulations of systems containing up to one million lattice sites to
  accurately determine the critical properties of these models. In strong
  contrast with earlier work, compelling quantitative evidence for the
  Kosterlitz--Thouless-like character of the phase transition is provided.
\end{abstract}


\begin{multicols*}{2}
One-dimensional scalar models with inverse-square potentials have held a
special position in the field of statistical mechanics for several decades.
The surprisingly rich behavior emerging from these conceptually simple systems
has attracted the attention of a variety of workers in the field. The work of
Ruelle~\cite{ruelle68} and Dyson~\cite{dyson69a} proved that pair interactions
decaying as $1/r^2$ (where $r$ denotes the distance between two local
variables) represent the boundary between models with and without a phase
transition at nonzero temperature; but it took until the 1980s before
Fr\"ohlich and Spencer~\cite{froehlich82} proved rigorously the existence of a
phase transition for the borderline case itself.  Meanwhile, early work by
Anderson and Yuval~\cite{anderson71} (who were motivated by the connection to
the Kondo problem) had suggested that this phase transition has a special
character, with an order parameter that exhibits both a discontinuity {\em
and\/} a superposed singularity at the critical point. This behavior later
gained widespread attention through the Kosterlitz--Thouless~(KT) transition in
the two-dimensional~(2D) $XY$ model~\cite{kt}. Indeed, the vortex-unbinding
scenario in the latter case has a direct analog in terms of logarithmically
interacting defects (domain walls) in the ``kink-gas'' representation of the
Ising chain considered in~\cite{anderson71}. This treatment in the spirit of
the renormalization-group~(RG) theory was extended in~\cite{kosterlitz76} and
most notably by Cardy~\cite{cardy81}, who demonstrated that it can be
generalized to large classes of discrete models.  In this sense, these
one-dimensional models can be considered as the simplest examples of the class
of ``topological'' phase transitions, which comprises phenomena like 2D
superfluidity and superconductivity, as well as dislocation-mediated melting of
crystals.

An exact lower bound on the critical order-parameter jump has been obtained by
Aizenman {\em et al.}~\cite{aizenman88} for Ising and Potts chains with $1/r^2$
interactions, and the existence of an intermediate ordered phase, where the
correlation function decays like a temperature-dependent power-law, has been
proven by Imbrie and Newman~\cite{imbrie88}.  Despite these rigorous results
the literature on these models contains various contradictory claims and
unconfirmed conjectures, while few of the RG-based predictions for the critical
properties have been verified by alternative methods: Thus (i)~the critical
temperature~$T_c$ has been estimated by a great variety of techniques, but
little consistency has been obtained until now, with estimates differing by up
to~30\%. Indeed, there is even a conjecture for $T_c$ of the $q$-state Potts
chain that is {\em independent\/} of~$q$~\cite{cannas96}, despite the fact that
this appears to contradict intuitive energy--entropy arguments. (ii) In
contrast, it has been claimed on the basis of numerical simulations that the
transition in the $q$-state Potts chain is continuous for $q \leq q_c$ and of
first order for $q>q_c$~\cite{bayong99}, although this in turn disagrees with
the KT-like transition (with $q$-dependent exponents) found by
Cardy~\cite{cardy81}.  (iii)~It has been conjectured that the power-law
exponent of the correlation function in the intermediate ordered phase is
directly related to the value of the order parameter~\cite{imbrie88}.  (iv)
Numerical simulations have been unable to conclude whether the specific heat
diverges at~$T_c$ or even whether it displays a maximum
at~$T_c$~\cite{bhatta81}.  (v) No numerical work for a one-dimensional system
has been able to reproduce one of the actual hallmarks of the KT universality
class, namely the predicted exponential divergence of the correlation length
and the susceptibility for $T \searrow T_c$.

It is the aim of the present work to settle these open questions in a
convincing way, although (ii) will be addressed only partially, by combining
finite-size scaling relations obtained from the RG predictions (which until now
have been ignored in nearly all numerical studies) with accurate numerical
simulations that span an unprecedented range in system sizes. Indeed, the
long-range nature of the interactions has posed a major hurdle for the
numerical study of these models: Whereas almost two decades ago Monte
Carlo~(MC) simulations could reach a system of $L=256$ lattice
sites~\cite{bhatta81}, the most recent work still could only attain
$L=900$~\cite{bayong99}. By using appropriate techniques, we have now been able
to study systems as large as $L=10^6$. This allows us not only to answer
questions intrinsic to the one-dimensional systems under study, but also to
obtain what we believe are the most accurate numerical results to date for a
system exhibiting a KT-like phase transition: Indeed, although the
2D $XY$ and Villain models have been examined in many numerical
studies, it has proven extremely difficult to obtain full consistency with the
RG predictions~\cite{janke}.  Because of the exponential divergence of the
correlation length~$\xi$, it is essential to cover a large range in {\em
linear\/} system sizes, which makes a one-dimensional system easier to study
than a 2D one. In our analysis, we start from the hypothesis that
the structure of the scaling functions is correctly described by RG theory, and
examine the consistency between our numerical results and the scaling
predictions to test the correctness of this hypothesis.

We have first focused attention on the Ising chain (or equivalently the $q=2$
Potts chain), described by the reduced Hamiltonian $\beta{\mathcal H} =
-K\sum_{\langle ij\rangle} s_i s_j/r_{ij}^2$, where the sum runs over all spin
pairs, $s_i=\pm 1$, and $r_{ij} = |{\mathbf r}_i-{\mathbf r}_j|$.  We have
carried out accurate MC simulations for chains of $L$ lattice sites,
with $10 \leq L \leq 10^6$, over a large temperature range. These results could
be obtained owing to the use of a cluster algorithm that suppresses critical
slowing down and has an efficiency that is {\em independent\/} of the
interaction range~\cite{lr-alg}.

\begin{figure}
\leavevmode
\centering
\epsfxsize\figurewidth
\epsfbox{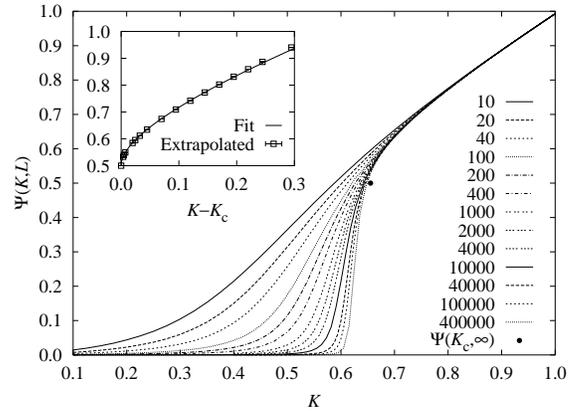}
\caption{The scaled order parameter $\Psi \equiv K\langle m^2\rangle$ of the
inverse-square Ising chain, as a function of the coupling~$K$. The inset shows
the estimates for $\lim_{L\to\infty}\Psi(K,L)$ for $K>K_c$, along with a fit of
the predicted singularity.}
\label{fig:psi}
\end{figure}

Although the critical coupling~$K_c$ is a nonuniversal quantity, the large
number of available estimates makes it interesting in its own right. In
addition, its precise value proves indispensable if one wants to address the
issues mentioned above. Since finite-size corrections decay only
logarithmically near a KT transition, the determination of~$K_c$ is notoriously
difficult, and a new approach had been found. To this end, we have relied on
the observation that the properly scaled order parameter $\Psi \equiv K \langle
m^2 \rangle$, with $m=L^{-1}\sum_{i=1}^{L} s_i$, is expected to exhibit a {\em
universal\/} jump at the critical point. Indeed, as mentioned, it has been
shown rigorously~\cite{aizenman88} that $\Psi(K>K_c) \geq \frac{1}{2}$ [whereas
$\Psi(K<K_c)=0$].  Furthermore, it was conjectured in~\cite{aizenman88} that
this inequality is saturated at $K=K_c$ (which {\em seems\/} to be implied by
the arguments of~\cite{anderson71}, but note the comments in~\cite{imbrie88}).
While it can be easily shown that $\Psi$ is invariant under a (real-space) RG
transformation of the kink-gas representation of the Ising chain~\cite{ktlong},
it appears not to have been explicitly realized that the critical jump $\Psi_c
\equiv \Psi(K_c)$ must be {\em universal}, in the sense that it will have the
same magnitude in every system with couplings that decay {\em asymptotically\/}
like $r^{-2}$.  We remark that also for the analogous phenomenon in two
dimensions (the celebrated universal jump in the superfluid density of a $^4$He
film) this universality was only realized~\cite{nelson77} {\em after\/} the
original work by Kosterlitz and Thouless~\cite{kt}. In addition to the jump
discontinuity, $\Psi(K>K_c)$ is predicted to exhibit a superposed singularity:
\begin{equation}
  \label{eq:tempsingularity}
  \Psi(K) = \Psi_c + a_1 (K-K_c)^{\tilde\nu} 
                   + a_2 (K-K_c)^{2\tilde\nu} + \ldots \;,
\end{equation}
with $\tilde\nu=\frac{1}{2}$ for the Ising chain; the ellipsis also includes
nonsingular correction terms.  In what follows, we demonstrate how to exploit
the scaling properties of $\Psi$ to accurately locate~$K_c$.
Figure~\ref{fig:psi} shows $\Psi(K,L)$ as a function of~$K$ for different
system sizes. While $\Psi$ is already sharpening up for relatively small~$L$,
it is only for very large systems that one observes the development of a
discontinuity and an additional singularity.

To proceed further, we recall the
lowest-order RG flow equations~\cite{anderson71,kosterlitz76} for the
coupling and the defect fugacity~$y$, namely
\begin{eqnarray}
  \label{eq:rg-eqs}
  dK/dl &=& -4Ky^2   \;,\\
  dy/dl &=& -y(2K-1) \;,
\end{eqnarray}
which (under the additional assumption that the system size scales with the
rescaling parameter $e^l$) allow us to derive the finite-size scaling behavior
of $\Psi(K,L)$ in the low-temperature regime.  Since this derivation is
somewhat involved~\cite{ktlong}, we only quote the final result here, namely
\begin{equation}
  \label{eq:fss-psi}
  \Psi(K,L) = \Psi(K,\infty)
  \left[1+b_1 L^{-2(\bar{\Psi}-1)} + b_2 L^{-4(\bar{\Psi}-1)} + \ldots \right]
  \;,
\end{equation}
where $\bar{\Psi} \equiv \Psi(K,\infty)/\Psi_c$.
Nonsingular corrections become important only when $\Psi > 2\Psi_c$.  The
appearance of $\Psi(K,\infty)$ both as an amplitude and as an exponent is
typical for a KT transition. A least-squares fit of this equation to
$\Psi(K,L)$ has been used to obtain $\Psi(K,\infty)$ for several values of $K$,
as shown in the inset in Fig.~\ref{fig:psi}. This also yielded $\Psi_c=0.49 \pm
0.02$.  Keeping $\Psi_c$ fixed at the predicted
value~$\frac{1}{2}$~\cite{aizenman88,ktlong} yielded $K_c=0.6552~(2)$.
Subsequently, Eq.~(\ref{eq:tempsingularity}) allowed us to determine
$\tilde\nu=0.52 \pm 0.03$. We have supplemented this low-temperature approach
with an analysis of data lying within the high-temperature finite-size regime,
where $L \ll \xi$ and $K<K_c$.  The algebraic corrections in~(\ref{eq:fss-psi})
now become powers of $1/\ln L$ and temperature-dependent terms take the form
$L/\xi$, with $\xi = \exp[B/(K-K_c)^{\tilde\nu}]$~\cite{ktlong}. We then find
$\Psi_c = 0.496~(3)$, in excellent agreement with the RG prediction, and
$K_c=0.6548~(14)$.  Keeping $\Psi_c$ now fixed at $\frac{1}{2}$ yields
$K_c=0.6555~(4)$, very close to the value resulting from the low-temperature
approach.

\begin{figure}
\leavevmode
\centering
\epsfxsize\figurewidth
\epsfbox{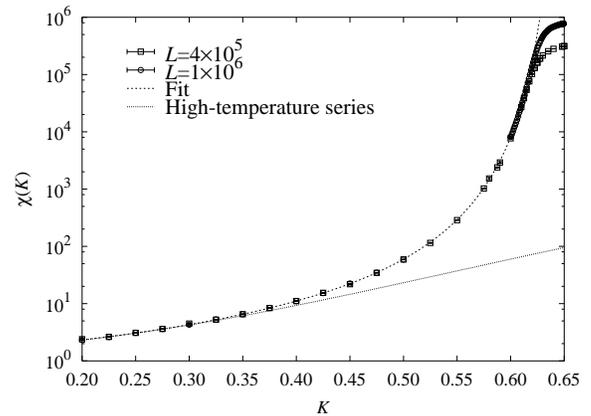}
\caption{The reduced magnetic susceptibility~$\chi = L^d \langle m^2\rangle$ of
the inverse-square Ising chain as a function of the coupling, along with a
(truncated) high-temperature series expansion~\protect\cite{matvienko85} and a
fit to the predicted asymptotic behavior
$\exp[B_0(K-K_c)^{-\tilde\nu}+B_2(K-K_c)^{\tilde\nu}]$.}
\label{fig:chi}
\end{figure}

Independent estimates for $K_c$ and~$\tilde\nu$ can be obtained from the
high-temperature susceptibility, which should display the same exponential
divergence as the correlation length. RG considerations regarding higher-order
contributions~\cite{amit76} suggest $\chi \sim \exp[B_0 (K-K_c)^{-\tilde\nu} +
B_2 (K-K_c)^{\tilde\nu}+{\mathcal O}((K-K_c)^{2\tilde\nu})]$. Numerical results
for $L=4\cdot 10^5$ and $L=10^6$ are shown in Fig.~\ref{fig:chi}, along with a
truncated high-temperature expansion of order ${\cal
O}(K^8)$~\cite{matvienko85} and a fit to the RG expression.  The agreement with
the series expansion is good, but breaks down for $K \gtrsim 0.3$. The RG
expression provides an excellent description of the MC data for $0.2 \leq K
\leq 0.61$, confirming the predicted exponential divergence over a large
temperature range. Owing to the covariance of the fit parameters, it is
difficult to determine them simultaneously: We obtained $\tilde\nu=0.54 \pm
0.03$ as our best estimate. However, keeping $\tilde\nu=\frac{1}{2}$ leads to
$K_c=0.6551~(6)$, which we view as a strong confirmation of our analysis
of~$\Psi$. The rounding of~$\chi$ close to~$K_c$ (see Fig.~\ref{fig:chi}) is
due to the finite system size. We estimate that the correlation length~$\xi$ is
$\gtrsim 10^5$ here, which may be compared with a maximum of $\xi \approx 140$
reached in recent $d=2$ simulations~\cite{janke}.

Subsequently, we carried out a similar analysis of $\Psi$ and $\chi$ for the
$q$-state Potts model with $q=3$, which has a Hamiltonian of the form $\beta
{\mathcal H}^{\rm Potts} = -K^{\rm Potts}\sum_{\langle ij\rangle} \delta_{s_i
s_j}/r_{ij}^2$, with $s_i=1,\ldots,q$. Cardy~\cite{cardy81} has found that, for
positive integer~$q$, this system can be described by RG equations that are
very similar to those for the $q=2$ (Ising) case, but yielding
$\tilde\nu=2/(q+2)$. This implies that the
entire analysis of the critical behavior can be carried through as
before~\cite{ktlong}: $\Psi$ should still exhibit a universal jump (of
magnitude~$1$ in units of~$K^{\rm Potts}$) with a superposed singularity as
in~(\ref{eq:tempsingularity}) and $\chi$ should again display an exponential
singularity.
The order parameter used for the determination of $\Psi$ and~$\chi$ is given by
$m^{\rm Potts} \equiv \max_{\alpha=1,\ldots,q} [\frac{q}{L}
\sum_{i=1}^{L}(\delta_{s_i\alpha}-1)] / (q-1)$, with $q>1$.  An analysis of
$\Psi$ for system sizes up to $L=10^6$ yielded $K_c=1.4105~(2)$,
$\Psi_c=1.02~(2)$, and $\tilde\nu(q=3)=0.41~(2)$. A least-squares fit of the
data for $\chi$ was consistent with an exponential divergence with
$\tilde\nu(q=3)=0.42~(2)$.  Not only is $\Psi_c$ in good agreement with the
predicted universal value, but to our knowledge this is also the very first
independent corroboration of the generalized RG scenario of
Ref.~\cite{cardy81}, which predicts that $\tilde\nu(q=3)=\frac{2}{5}$. As
expected from entropic considerations, $K_c^{\rm Potts}(q=3)$ is somewhat
higher than $K_c^{\rm Potts}(q=2)=1.3104~(4)$, which in turn refutes two
conjectures based on a different real-space RG analysis~\cite{cannas96}, namely
that (i)~$K_c^{\rm Potts}$ would be {\em independent\/} of~$q$ and
(ii)~$K_c^{\rm Potts}$ would be exactly given by $12/\pi^2=1.21585\ldots$ We
note that our results are in strong contrast with earlier numerical work: Both
a transfer-matrix study and a very recent MC study for $q=2$~\cite{glumac93}
and $q=3$~\cite{bayong99,glumac93} concluded that the correlation length
diverges {\em algebraically\/} (rather than exponentially) with a $q$-dependent
power-law exponent.

\begin{figure}
\leavevmode
\centering
\epsfxsize\figurewidth
\epsfbox{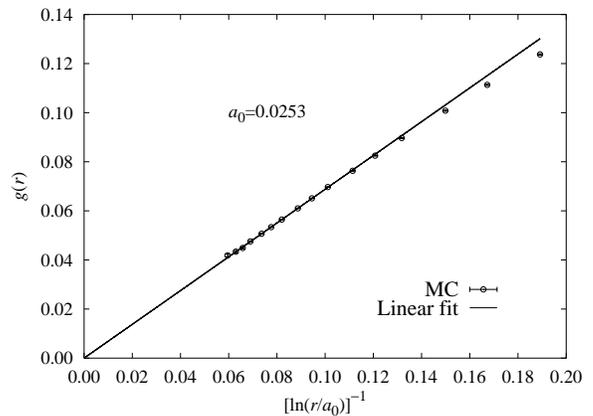}
\caption{The critical correlation function $g(r)$ of the inverse-square Ising
chain, for $5 \leq r \leq 5\cdot 10^5$. The logarithmic decay can be followed
for at least three decades in~$r$.}
\label{fig:gkc}
\end{figure}

\begin{figure}
\leavevmode
\centering
\epsfxsize\figurewidth
\epsfbox{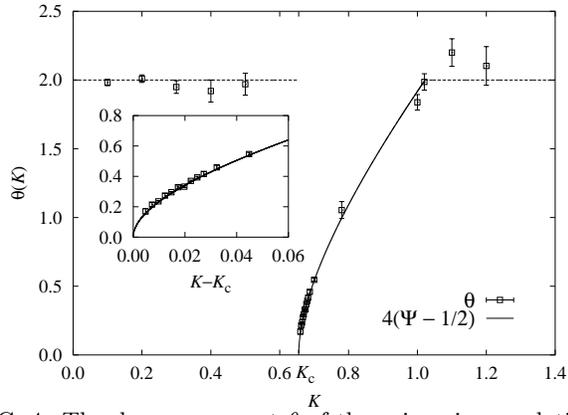}
\caption{The decay exponent~$\theta$ of the spin-spin correlation
function~$g(r)$ of the inverse-square Ising chain, as a function of the
coupling.}
\label{fig:theta}
\end{figure}

We now return to some interesting aspects of the Ising chain. First, it has
been concluded in Ref.~\cite{bhatta81} that the critical spin-spin correlation
function should decay logarithmically, $g(r) = \langle s_0 s_r \rangle -
\langle s_0 \rangle ^2 \sim 1/\ln r$, where now $\langle s_0\rangle \neq 0$,
unlike the situation for a typical critical phase transition.  As illustrated
in Fig.~\ref{fig:gkc}, we have been able to corroborate this behavior over a
considerable range in~$r$. In the high-temperature phase, the decay of $g(r)$
is bounded by the spin-spin interaction, but for $K>K_c$ it is expected that
$g(r) \sim r^{-\theta}$, with $\theta = 4\sqrt{(K-K_c)/K_c}$~\cite{bhatta81}.
We note that this behavior would be consistent with the conjecture of
Ref.~\cite{imbrie88}, {\em viz.\/}\ that $\theta=\min[4(\Psi-\frac{1}{2}),2]$
for $K>K_c$. We have determined $\theta$ for a large number of couplings, see
Fig.~\ref{fig:theta}, and have subsequently fitted these to a power series in
$\sqrt{K-K_c}$, finding $\theta = (1.81 \pm 0.06) \sqrt{(K-K_c)/K_c}$. This
agrees with $4a_1 \sqrt{K_c} = 1.73~(4)$, where $a_1$ is defined via
Eq.~(\ref{eq:tempsingularity}), corroborating the conjectured relation
with~$\Psi$.  However, the prefactor is approximately half as large as
predicted in Ref.~\cite{bhatta81}.  Farther away from $K_c$ it becomes
increasingly difficult to obtain accurate results for $\theta(K)$, owing to the
rapid decay of~$g(r)$. However, since our results corroborate the presumed
relation between $\theta$ and $\Psi$, we exploit this relation to determine two
special values of~$K$, namely the coupling strength~$K_1$ for which $\theta=1$
and the smallest coupling~$K_2>K_c$ for which $\theta=2$.  $K_1$ has the
special property that the magnetic susceptibility will diverge over the full
coupling range $K_c \leq K \leq K_1$, as follows from the
fluctuation--dissipation theorem (and as is actually confirmed by the
numerical results~\cite{ktlong}). We find $K_1=0.78~(1)$ and $K_2=1.02~(3)$.
Thus, the RG prediction $K_1=\frac{17}{16}K_c \approx 0.696$~\cite{bhatta81} is
too low, due to the difference in the prefactor for $\theta(K)$.  $K_2$ marks
the end of the regime where $g(r)$ displays a leading temperature-dependent
exponent.

\begin{figure}
\leavevmode
\centering
\epsfxsize\figurewidth
\epsfbox{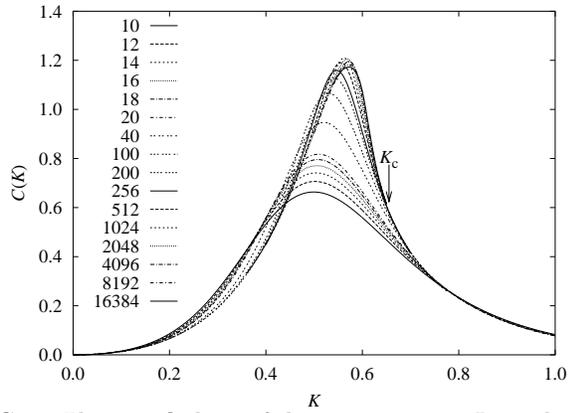}
\caption{The specific heat of the inverse-square
Ising chain, for a range of system sizes.  Note that the maximum reaches its
thermodynamic limit nonmonotonically.}
\label{fig:cv}
\end{figure}

Finally, we consider the specific heat~$C$, which is calculated from the
fluctuations in the internal energy. The efficiency of the original MC
algorithm~\cite{lr-alg} is greatly reduced when this energy has to be
calculated (although recently this problem has largely been
overcome~\cite{krech00}) and hence only much smaller systems could be accessed.
Nevertheless, this turns out not to be a limiting factor, since $C$ has already
converged within the range of accessible system sizes: Fig.~\ref{fig:cv} shows
unambiguously that $C$ is a nondivergent quantity which takes its maximum at
a temperature that lies approximately 14\% above the critical temperature.

In summary, we have provided compelling evidence that the inverse-square Ising
and $q=3$ Potts chain display the one-dimensional analog of a
Kosterlitz--Thouless transition, in disagreement with earlier numerical
studies, but in accurate quantitative agreement with renormalization-group
predictions.  The (exponentially diverging) correlation length could be
examined over more than five orders of magnitude, making this the most
tractable KT-like transition.

It is a great pleasure to acknowledge stimulating remarks by Prof.\ Kurt Binder
and helpful comments by Prof.\ Michael E. Fisher. E.L. gratefully acknowledges
the hospitality of the Condensed Matter Theory Group at the Johannes
Gutenberg-Universit\"at Mainz, where most of this work has been carried out,
and support from the National Science Foundation (through Grant No.\ CHE
99-81772 to M.E. Fisher) and from the Department of Energy, Office of Basic
Energy Sciences (through Grant No.\ DE-FG02-98ER14858 to A.Z.
Panagiotopoulos).

\end{multicols*}

\end{document}